# EEG-DIF: Early Warning of Epileptic Seizures through Generative Diffusion Model-based Multi-channel EEG Signals Forecasting

Zekun Jiang[1,2,†], Wei Dai[2,†], Qu Wei[2], Ziyuan Qin[2], Kang Li[2, *] and Le Zhang[1,*]

[1]College of Computer Science, Sichuan University, Chengdu, China

[2]West China Biomedical Big Data Center, West China Hospital, Sichuan University, Chengdu, China

*Corresponding Authors: zhangle06@scu.edu.cn, likang@wchscu.cn

## ABSTRACT

Multi-channel EEG signals are commonly used for diagnosis and assessment of diseases such as epilepsy. Currently, various EEG diagnostic algorithms based on deep learning have been developed. However, most research efforts focus solely on diagnosing and classifying current signal data, but not consider the prediction of future trends for early warning.

Additionally, since multi-channel EEG can be essentially regarded as the spatio-temporal signal data received by detectors at different locations in the brain, how to construct spatio-temporal information representations of EEG signals to facilitate future trend prediction for multi-channel EEG becomes an important problem. This study proposes a multi-signal prediction algorithm based on generative diffusion models (EEG-DIF), which transforms the multi-signal forecasting task into an image completion task, allowing for comprehensive representation and learning of the spatio-temporal correlations and future developmental patterns of multi-channel EEG signals.

Here, we employ a publicly available epilepsy EEG dataset to construct and validate the EEG-DIF. The results demonstrate that our method can accurately predict future trends for multi-channel EEG signals simultaneously. Furthermore, the early warning accuracy for epilepsy seizures based on the generated EEG data reaches 0.89. In general, EEG-DIF provides a novel approach for characterizing multi-channel EEG signals and an innovative early warning algorithm for epilepsy seizures, aiding in optimizing and enhancing the clinical diagnosis process.

## 1 INTRODUCTION

Epilepsy, or a seizure disorder, is a chronic disease that affects around 50 million people of all ages globally, ranking it among the most common neurological disorders in the world [1]. Epileptic seizures are typically sudden and characterized by abrupt bursts of abnormal electrical activity in the brain, potentially leading to numerous negative consequences, such as injuries resulting from sudden loss of consciousness and the development of long-term anxiety and depression. These effects can be both immediate and lasting, seriously compromising the physical and mental health as well as the quality of life of patients, and may even lead to death [2]. Besides surgical treatment, antiepileptic drugs (AEDs) can control seizures to a certain extent, but they still have limitations. According to the World Health Organization (WHO), these AEDs fail in about 30% of patients suffering from seizure disorders [3, 4]. For this reason, predicting and providing early warnings of seizures is becoming crucial for implementing early interventions to protect patients from severe consequences.

Electroencephalography (EEG) is considered as one of the most effective and consistent predictors to monitor epileptic seizures [5]. It typically involves placing multiple electrodes on the patient's scalp to collect real-time electrical activity signals from various brain regions, which are then used to detect abnormal activities. It has been demonstrated that EEG features, such as rapid spiking waves, can serve as indicators of epileptic seizures [6]; therefore, in theory, multi-channel EEG signals can provide insights on seizure early warning.

Artificial intelligence (AI) technology, especially advanced deep learning methods, has brought technological innovations to the intelligent diagnosis and prediction of EEG signals [7, 8]. However, there are still several technical challenges in the field of EEG analysis that affect the clinical implementation of EEG-based deep learning models.

Firstly, multi-channel EEG signals can essentially be considered as multi-source spatiotemporal signal data received by multiple detectors located in different brain regions, encompassing both temporal and spatial information. Current research methods often focus only on temporal correlations while neglecting spatial connections [9, 10]. Therefore, how to comprehensively characterize the spatio-temporal relationships for multi-channel EEG signals to build up a superior deep learning model is becoming a critical challenge that needs to be addressed in the current field.

Secondly, classical deep learning methods used for EEG signals forecasting, such as the commonly used LSTM algorithm, typically predict the future pattern of only one channel signal per model [11-13]. Nonetheless, it indicates that early warning for multi-channel EEG signals would request simultaneous inference computations by over a dozen deep learning models, which is not conducive to clinical practice. Currently, how to develop such an algorithm that can predict future trends for multiple channels simultaneously with a single model is becoming a pressing issue that needs to be addressed.

Lastly, most current EEG-based deep learning studies build up classification models based solely on the currently acquired EEG signals [14, 15], lacking experimental evidence for classification and diagnosis over a future period. However, early warning of epileptic seizures requests developing such a diagnostic model for EEG signals that can predict future trends of multi-channel EEG signals.

Recently, the emergence and development of generative AI algorithms such as GANs [16, 17] and Diffusion models [18, 19] provide powerful technical approaches for multimodal data processing and future prediction, helping to address challenges that classical deep learning algorithms are unable to solve. Therefore, this study aims to develop a novel generative model-based spatio-temporal representation, future trend forecasting, and early warning algorithm for multi-channel EEG signals to solve the aforementioned problems. We demonstrated that our novel approach can accurately predict future trends of multi-channel EEG signals and obtain early diagnosis of epileptic seizures after evaluating our method on a public epilepsy EEG dataset.

## 2　RELATED WORK

### 2.1　Spatio-temporal representation of EEG signals

EEG signals are formed by real-time detection of electrical signals from detectors located in different brain regions. In addition to temporal characteristics, they also have spatial correlations and can be considered as spatio-temporal data. Therefore, how to represent the spatio-temporal relationships of EEG signals is crucial for us to develop deep learning based predictive models [20-26].

Most current studies are exploring the spatio-temporal representation learning of EEG signals [27-32]. Generally, there are two main approaches for representation learning: one is converting EEG signals into two-dimensional or three-dimensional images, then modeling them using CNN or CNN-LSTM networks [27-29]; the other is constructing a graph structure between different EEG signals and then developing Graph Neural Network models [31, 32]. These spatio-temporal representation learning algorithms mainly rely on the transformed data forms and do not focus much on the spatio-temporal correlations between different channels and time points of EEG signals during the learning process. Therefore, it is essential to develop a novel algorithm for spatio-temporal representation learning using EEG signals.

### 2.2　Multi-channel EEG signals forecasting

Sequence signal forecasting is an important research direction, but most previous studies could only predict a single channel signal, such as LSTM and other RNN algorithms [33-35]. The emergence of generative AI algorithms, such as GANs and Diffusion models, has made the generation and forecasting of multi-channel signals possible. However, current generative AI research based on EEG mainly focuses on data augmentation tasks. For example, Xu et al. [36] have developed a GAN-based multi-channel synthetic algorithm to enhance the EEG prediction for epileptic seizures, mainly overcome the data insufficiency. Also, Shu et al. [18] have introduced a promising method called DiffEEG for EEG data augmentation by using a diffusion model. Their work compares pseudo-data generated by DiffEEG with that from classical augmentation techniques, through MLP, CNN, and Transformer classifiers, and the same classifiers demonstrate superior results with DiffEEG generated data.

It is evident that current related works primarily emphasize improving multi-channel EEG data augmentation techniques to enhance model performance, while also demonstrating the potential of using generative AI approaches for modeling. However, these studies lack research on the generation and prediction of future trends in multi-channel EEG signals. Since there is a strong correlation between multi-channel EEG signals, it is essential to develop a model capable of simultaneously predicting various interrelated signals. Moreover, because the task of multi-channel signal forecasting presents greater difficulty and challenges, it urgently requests the development of novel algorithms to address these issues.

### 2.3　Early warning of seizures using EEG signals

Applications of seizure early warning using EEG signals represent a relatively new field of research, while more research has been conducted on seizure detection using EEG data. For instance, Nallur et al. [15] developed an early warning system that combines the African Vultures Optimization Algorithm for feature selection with channel and spatial attention. This system is trained with selected preictal data for future seizure prediction. Also, Xie et al. [32]. explored seizure brain connectivity using a spatial-temporal graph neural network (GNN) method, demonstrating the potential of GNN-extracted features in seizure detection and prediction when processed with MLP layers. Additionally, Zhou et al. [37] proposed GlepNet method, which combines a temporal convolutional layer with a multi-head attention mechanism to capture spatial-temporal information within preictal EEG signals, showing promising results across various seizure datasets. Furthermore, Li et al. introduced GGN and Transformer model for epileptic seizure detection [31] and Cho et al. proved potential of treating EEG signals as multi-channel images [14].

In summary, most of the studies we reviewed focus on the current epileptic seizures' detection, but they remain a significant gap in providing early warnings for seizures. Since the few existing works [32] on future prediction of seizures have not been tested on future-generated signal data, it lacks experimental evidence.

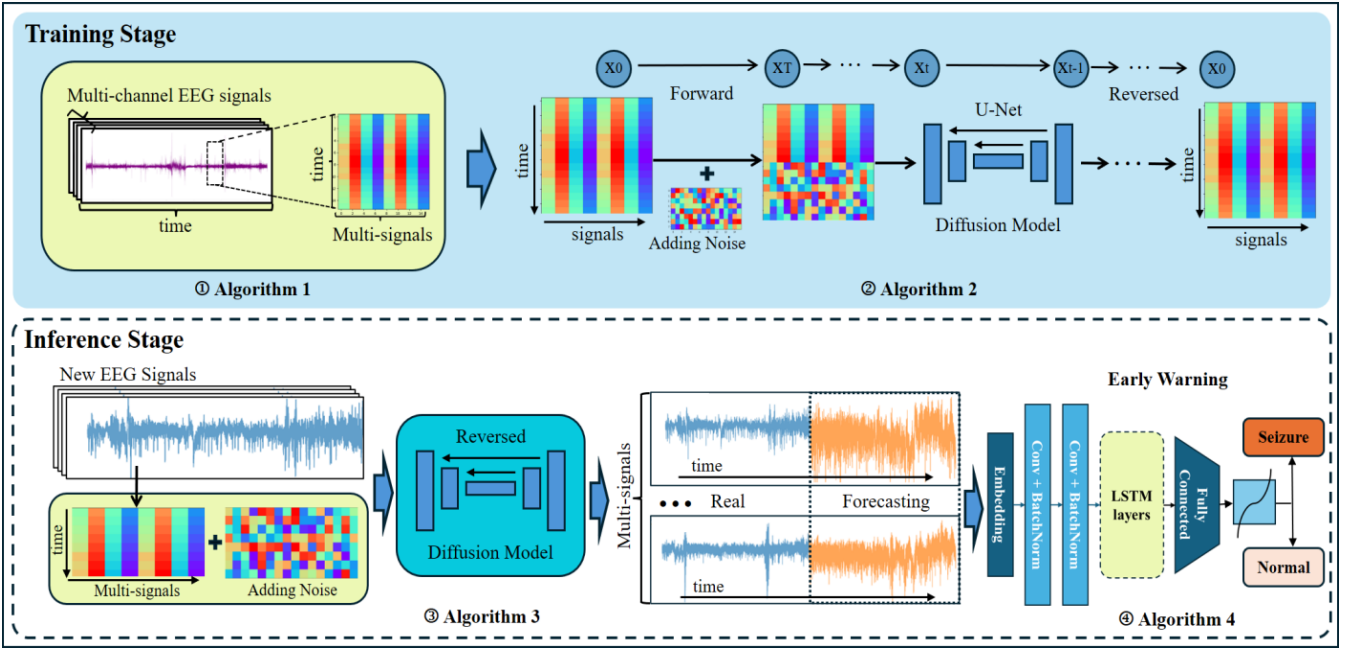

Figure 1: EEG-DIF algorithm framework. In the training stage, translating multi-signal forecasting into image completion is implemented by Algorithm 1, then EEG-DIF model training is implemented by using DDIM (Algorithm 2). In the inference stage, EEG-DIF model is used for new EEG signals to generate future signals by Algorithm 3, and early warning of seizure is retaliated by using CNN-LSTM based on the generated data by Algorithm 4.

Therefore, it is very challenge for us to combine future multi-channel generated signals to construct early warning diagnostic models.

### 2.4 Key contributions

To address the aforementioned technical challenges, we propose three innovative solutions and develop a novel early warning algorithm for epileptic seizures. This algorithm, named EEG-DIF, leverages a generative diffusion model to forecast multi-channel EEG signals. Our main contributions are listed as follows:

1. We present a novel view by transforming multi-signal forecasting into image completion tasks, which helps us to represent and mine the spatio-temporal relationships for multi-channel EEG signals, enhancing signal generation and prediction tasks.

2. We propose a novel and efficient multi-signal forecasting algorithm by using diffusion model to simultaneously predict future trends for multi-channel EEG signals.

3. We integrate a CNN-LSTM classifier into the backend of EEG-DIF and develop an early warning diagnostic model for epileptic seizures based on the generated EEG signals, providing accurate early seizure predictions.

## 3 METHODS AND DATA

Before introducing the specific solutions, we first present the workflow of the EEG-DIF algorithm by **Figure 1**. Briefly, we convert multi-channel signals into two-dimensional images and train an image completion model [38, 39] based on the generative diffusion model to fully characterize the spatio-temporal relationships of multi-channel EEG signals. When new multi-channel signal data is received, it is converted into two-dimensional images and merged with noise tensors of the same dimensions along the time axis. The trained diffusion model is then used to denoise the noisy parts of the images, achieving future trend prediction for multiple signals. Based on the generated future signal data, we construct an epileptic seizure diagnostic model to have accurate early seizure warnings.

Algorithms 1-4 represent the four main stages of the EEG-DIF algorithm during the training and inference phases. Each of these stages will be detailed in the following subsections.

### 3.1 Translating multi-signal forecasting into image completion

The foundation of EEG-DIF involves translating the multi-signal prediction task into an image completion task, which converts a set of one-dimensional signals into a two-dimensional waveform image as **Algorithm 1** of **Figure 1**.

$$S'_{i,j} = \frac{S_{i,j} - \min(S_i)}{\max(S_i) - \min(S_i)} \quad (1)$$

First, we use Equation 1 to rescale each original signal channel $S_i$ by applying min-max normalization[40-46], bringing it into the range between 0 and 1, resulting in the normalized multi-channel EEG signal data $S'_{i,j}$. Here, $i$ represents the signal channel and $j$ represents the time point.

$$I(j,i) = S'_{i,j} \qquad (2)$$

Then, we use Equation 2 to convert the multi-channel signal data $S'_{i,j}$ into a two-dimensional tensor image $I(j,i)$ with dimensions of time × signals. All signals are simultaneously sampled at the same rate, making it feasible to combine them into a two-dimensional signal image in this way, sharing a common temporal dimension $j$.

## 3.2 EEG-DIF model training

Based on the obtained signal image data $I(j,i)$, we train the EEG-DIF model by **Algorithm 2** of **Figure 1**.

The EEG-DIF algorithm employs a Denoising Diffusion Implicit Model (DDIM) [47] with backbone of U-Net [48]. The training of EEG-DIF follows the generative diffusion model training routine, with several modifications to the training loop.

$$q(x_t | x_{t-1}, x_0) = \frac{q_\sigma(x_{t-1} | x_t, x_0) q_\sigma(x_t | x_0)}{q_\sigma(x_{t-1} | x_0)} \qquad (3)$$

In the forward/noising process, the noisy image $x_t$ is obtained from the original signal image $x_0$ by using Equation 3. Rather than adding Gaussian noise [49-52] to the entire image, EEG-DIF applies noise addition only to the bottom half of the signal image, which essentially means adding noise starting from a fixed point on the timeline. This non-Markovian process is formulated using Bayes' rule [47], where each intermediate variable $x_t$ is dependent on the variable $x_{t-1}$ at the next time step and the data $x_0$ at original time point, with the variability of the process controlled by the parameter $\sigma$.

$$x_{t-1} = \sqrt{\alpha_{t-1}} \left( \frac{x_t - \sqrt{1-\alpha_t} \varepsilon^{(t)} \theta(x_t)}{\sqrt{\alpha_t}} \right) + \sqrt{1-\alpha_{t-1} - \sigma_t^2} \cdot \varepsilon_\theta^{(t)}(x_t) + \sigma_t \varepsilon_t \qquad (4)$$

Subsequently, during the reversed/denoising process, the EEG-DIF model learns and fits Equation 4, gradually reconstructing the data from its noisy state $x_t$ to an intermediate state $x_{t-1}$, and finally restoring it to the original data $x_0$.

## 3.3 EEG-DIF model inference

Based on the trained EEG-DIF model, EEG-DIF inference is implemented upon receiving new multi-channel EEG signals, as shown in **Algorithm 3** of **Figure 1**.

Specifically, we first use Equations 1-2 to convert the incoming signal data $S$ into normalized signal data $S'$ and a signal image $I$. Then, based on Equation 3, we concatenate the signal image with the generated Gaussian noise image to obtain the original noisy image $x_t$, where the noise is generated according to the shape of the signal image tensor along the time dimension.

Subsequently, $x_t$ is input into the EEG-DIF model, which uses the learned rules to apply Equation 4 to reconstruct the noisy region, obtaining a new image $x_0$. Based on this inference method, the gradually noised image can be continuously reconstructed and completed to obtain a generated image $I_{new}$ with sufficient time length.

Finally, using the corresponding relationship described by Equation 2, we can obtain the new multi-channel EEG signal data $S'_{new}$ from the generated image $I_{new}$. Using the stored min-max values from Equation 1, each channel signal of $S'_{new}$ is restored to its original scale to have the final EEG signals $S_{new}$.

As illustrated in **Algorithm 3** of **Figure 1**, the real-time received EEG signals can be used to generate multi-channel signal data to reflect future trends by implementing EEG-DIF inference. Additionally, by setting appropriate training and inference parameters, EEG-DIF theoretically allows for the prediction of signals of any number of channels and any length of time.

## 3.4 Early warning of epileptic seizures

Based on the aforementioned operations, future signal data can be generated from the currently received signals. As shown in **Algorithm 4** of **Figure 1**, the generated signals are used as inputs for an epileptic seizure detection algorithm, forming an epilepsy early-warning system.

This study integrates a CNN-LSTM classifier [53, 54] into EEG-DIF framework for epileptic seizure diagnosis. The CNN-LSTM model is composed of two convolutional layers and an LSTM network.

$$Y = CNNLSTM(S_{new}) \qquad (5)$$

Here, by using Equation 5, the generated EEG signals $S_{new}$ is input into the CNN-LSTM model, which outputs whether an epileptic seizure will occur, thereby achieving early warning of epileptic seizures.

## 3.5 Datasets

In the study, to comprehensively evaluate the effectiveness of our method, we used an open Siena Scalp EEG Database [55, 56] from PhysioNet (https://physionet.org/content/siena-scalp-eeg/1.0.0/) to construct and test the EEG-DIF generation and early-warning model. This dataset is acquired from the Unit of Neurology and Neurophysiology of the University of Siena, containing EEG data from 14 patients—9 males and 5 females across various age ranges with a sampling rate of 512Hz. Each patient includes multiple EEG electrode signals and one or two EKG signals, with electrodes arranged according to the international 10-20 system. Patient information includes epilepsy classification according to the criteria of the International League Against Epilepsy, the number of seizures, the number of EEG channels, and the total recording time in minutes. Overall, the dataset comprises approximately 128 recording hours with 47 seizures.

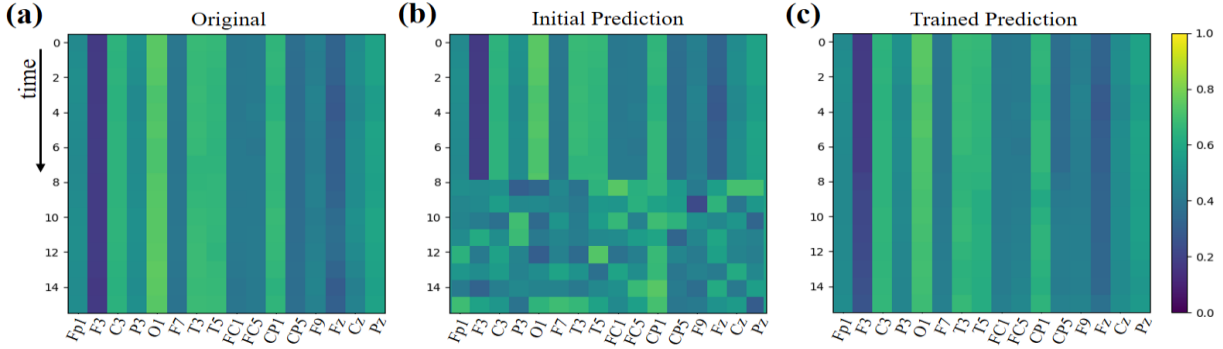

**Figure 2: Heatmaps for demonstrating spatio-temporal representation performance. Figures (a) shows the original signal image. Figure (b) and (c) depict the model's predicted signal image with initial weights and trained weights respectively. The vertical axis of the signal image represents the time axis from top to bottom, and the horizontal axis represents different EEG electrode channels with their spatial positions.**

To standardize the number of EEG electrodes and facilitate experimental validation, we selected 16 EEG channels from different brain regions that are most commonly used for epileptic seizure detection based on research experience [4, 56]. These channels are arranged in spatio-temporal order as follows: Fp1, F3, C3, P3, O1, F7, T3, T5, FC1, FC5, CP1, CP5, F9, Fz, Cz, and Pz.

Additionally, to develop the epileptic seizure classification model, we perform sliding window segmentation on the selected signal data and labeled whether a seizure occurred. The time window is approximately 30 seconds. Ultimately, we made 400 training samples for the classification modeling. The trained diagnostic model will be validated on the signal data generated by the EEG-DIF model.

### 3.6 Evaluation metrics

To evaluate the prediction accuracy and goodness of fit for the multi-signal forecasting algorithm, we employ the mean absolute error (MAE), mean square error (MSE), root mean square error (RMSE), and the coefficient of determination ($R^2$) to quantify the predictive error and the proportion of variance explained.

MAE, MSE, and RMSE are used to evaluate the degree of regression fit, with smaller values indicating better performance; $R^2$ is used to explain the proportion of variance, and typically, the closer the value is to 1, the better.

We employ t-test ($P<0.05$) to indicate a significant difference between different regression performances[57-59]. Additionally, we conduct qualitative analysis and visualization in the form of heatmaps and signal curve plots, providing a comprehensive understanding of our model.

For the evaluation of classification diagnostics, we conduct a quantitative analysis using the Receiver Operating Characteristic (ROC) curves and the Area Under the ROC Curve (AUC) values, as well as accuracy, precision, and F1 score metrics. The statistical differences between different algorithm models are assessed using the DeLong test [60, 61], with $P<0.05$ showing significant difference.

**Table 1: Comparison of the average prediction performance of the trained model versus the initialized model for multi-channel EEG.**

| Model | MAE | MSE | RMSE | $R^2$ |
|---|---|---|---|---|
| Initial | 65.278 | 7105.765 | 80.745 | 0.003 |
| Trained | 7.875 | 114.923 | 10.151 | 0.765 |

## 4 RESULTS

### 4.1 Spatio-temporal representation performance of EEG-DIF

EEG-DIF performs spatio-temporal representation of EEG signals by transforming signal prediction into image completion. **Figure 2** shows an example of a signal image to demonstrate the spatio-temporal representation and performance of the EEG-DIF algorithm for multi-channel signal forecasting. Here, a 16×16 signal image is constructed, and the lower half of the signal image is noised and restored to train the image completion model. Figure (a) is the input original signal image, while the Figure (b) and (c) show the reconstruction effect of the model with initial weights and trained weights after noise is added, respectively. Here, the trained model Figure (c) can better reconstruct the trend of the signal image than the initial model Figure (b).

Furthermore, **Table 1** presents a statistically significant improvement in spatio-temporal representation and prediction of EEG signals using our method (P<0.001), as demonstrated by quantitative metrics and statistical analysis.

Table 2: Comparison of the forecasting performance of the EEG-DIF and traditional LSTM algorithm across 16 EEG channels. MAE, MSE, and RMSE are used to evaluate the degree of regression fit, with smaller values indicating better performance; $R^2$ is used to explain the proportion of variance, with values closer to 1 signifying better performance.

| EEG Channel | EEG-DIF (one model for 16 signals forecasting) | | | | LSTM (one model for one signal forecasting) | | | |
|---|---|---|---|---|---|---|---|---|
| | MAE | MSE | RMSE | $R^2$ | MAE | MSE | RMSE | $R^2$ |
| Fp1 | 9.651 | 189.691 | 13.772 | 0.671 | 10.022 | 808.370 | 28.432 | 0.153 |
| F3 | 6.639 | 77.385 | 8.796 | 0.449 | 9.879 | 367.277 | 19.164 | 0.044 |
| C3 | 6.909 | 94.588 | 9.725 | 0.827 | 7.305 | 129.104 | 11.362 | 0.584 |
| P3 | 9.013 | 138.652 | 11.775 | 0.860 | 10.180 | 162.478 | 12.747 | 0.421 |
| O1 | 9.909 | 171.979 | 13.114 | 0.807 | 9.757 | 249.256 | 18.217 | 0.028 |
| F7 | 3.705 | 24.800 | 4.979 | 0.484 | 5.718 | 151.209 | 8.156 | 0.021 |
| T3 | 5.961 | 57.436 | 7.578 | 0.909 | 8.013 | 100.627 | 10.031 | 0.362 |
| T5 | 9.327 | 154.134 | 12.415 | 0.850 | 10.890 | 353.332 | 22.382 | 0.268 |
| FC1 | 6.321 | 60.928 | 7.805 | 0.869 | 7.846 | 173.423 | 11.568 | 0.401 |
| FC5 | 5.906 | 60.286 | 7.764 | 0.897 | 6.762 | 171.625 | 9.465 | 0.330 |
| CP1 | 7.035 | 77.765 | 8.818 | 0.892 | 8.433 | 264.812 | 18.051 | 0.514 |
| CP5 | 7.238 | 88.435 | 9.403 | 0.891 | 7.744 | 163.628 | 10.976 | 0.351 |
| F9 | 9.995 | 156.836 | 12.523 | 0.886 | 10.824 | 451.294 | 22.300 | 0.146 |
| Fz | 3.215 | 16.439 | 4.054 | 0.547 | 3.964 | 245.496 | 14.954 | 0.121 |
| Cz | 9.610 | 133.823 | 11.568 | 0.818 | 11.528 | 208.404 | 14.436 | 0.296 |
| Pz | 15.566 | 335.590 | 18.319 | 0.772 | 16.646 | 734.730 | 24.850 | 0.160 |
| Average | 7.875 | 114.923 | 10.151 | 0.777 | 9.094 | 295.942 | 16.068 | 0.263 |

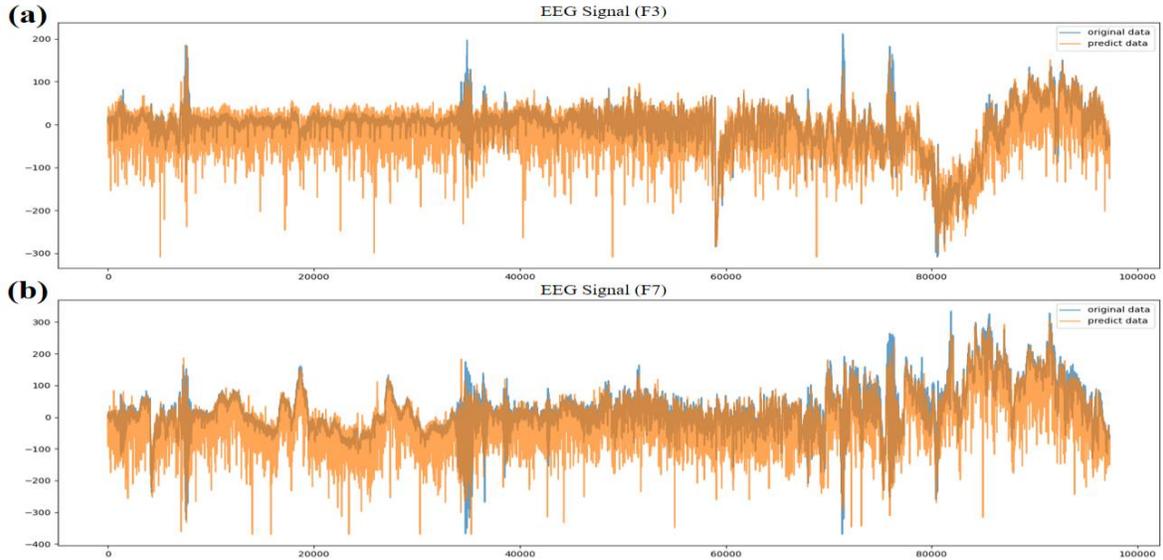

Figure 3: Prediction signal curves of the EEG-DIF model when encountering new data. Figure (a) and (b) depict the F3 and F7 EEG channel signals of the test patient, with the blue curve representing the original signal data and the orange curve representing the model generated signal data. The prediction time length is 190 seconds, generating a total of 97,280 time points.

### 4.2 Forecasting performance for Multi-channel EEG signals

The trained EEG-DIF model can simultaneously predict the future trends of multiple EEG signal channels. To evaluate its forecasting performance on each channel, **Table 2** presents the quantitative evaluation metrics of EEG-DIF on 16 EEG channels from test patients and compares them with the classical LSTM method, demonstrating the advance of our approach for multi-channel signal forecasting.

By analyzing the multiple evaluation metrics in **Table 2**, it can be observed that the prediction performance of a single EEG-DIF model exceeds that of 16 individual LSTM models across all 16 EEG channel signals. Additionally, the average prediction performance of EEG-DIF is significantly better than the average performance of the 16 LSTM models (P<0.05).

Meanwhile, we find that there is some variability in the prediction performance of EEG-DIF across different EEG signal channels. Specifically, the $R^2$ metric is relatively less than other channel signals for the F3 and F7 channels, being less than 0.5.

To further investigate these relatively poor prediction results, **Figure 3** describes the EEG-DIF prediction signal curves for F3 and F7 channels. **Figure 3** shows several regions where amplitudes will increase or vary, which affects the evaluation metrics. However, our model can still learn the trend of the development patterns of the signals, since the predicted signals (orange curves) show a considerable trend similarity to the real signals (blue curves), especially in accurately simulating some significant peaks.

## 4.3 Early warning performance for epileptic seizures

To evaluate the feasibility and potential of our method for early epilepsy diagnosis prediction, we construct the CNN-LSTM diagnostic model based on the training dataset (400 samples) and assess its diagnostic performance on 100 signal data generated by EEG-DIF prediction. Additionally, we compare its performance with the classical classification algorithms, including Logistic Regression (LR), XGBoost, Random Forest (RF), and Support Vector Machine (SVM).

**Figure 4** presents the ROC curves for various classification diagnostic models. Our epilepsy diagnosis model achieves an AUC of 0.89, significantly outperforming other classifiers (P<0.05). Detailed classification evaluation metrics are provided in **Table 3**, the CNN-LSTM model has an accuracy of 0.89, a precision of 0.93, and a F1-Score of 0.88, showing the best diagnostic advantages in accuracy, precision, and F1-Score.

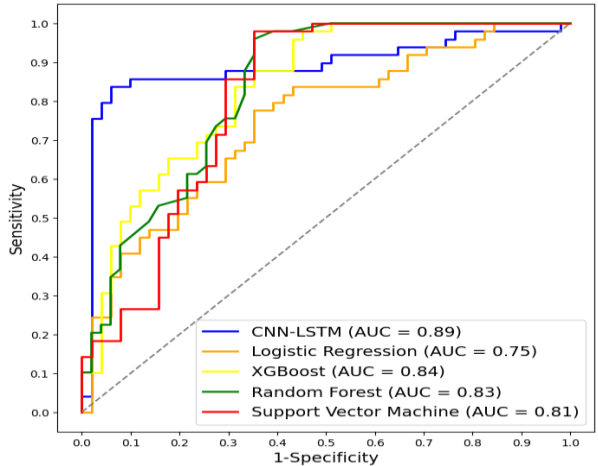

**Figure 4: ROC curves for comparing the diagnostic performance of CNN-LSTM and other classifiers.**

**Table 3: Model performance of different diagnostic algorithms. The value with bold indicates the best performance.**

| Models | AUC | Accuracy | Precision | F1-Score |
|---|---|---|---|---|
| **CNN-LSTM** | **0.89** | **0.89** | **0.93** | **0.88** |
| LR | 0.75 | 0.71 | 0.68 | 0.72 |
| XGBoost | 0.84 | 0.76 | 0.68 | 0.80 |
| RF | 0.83 | 0.80 | 0.72 | 0.82 |
| SVM | 0.81 | 0.81 | 0.73 | 0.83 |

## 5 DISCUSSION AND CONCLUSION

Here, we propose a novel early warning algorithm and system for epileptic seizures based on spatio-temporal representation and future forecasting of multi-channel EEG signals, named EEG-DIF. This algorithm uses a Diffusion model to predict future trend for any number of channel signals by completing signal image inpainting. Simultaneously, it implements early diagnostic classification based on the generated future signal data. The EEG-DIF demonstrates excellent signal forecasting and epileptic seizure diagnosis performance on public epilepsy datasets.

Extensive experiments demonstrate that EEG-DIF addresses key technical issues in the field from the following three aspects:

Firstly, as shown in **Figure 2** and **Table 1**, converting multi-channel signal forecasting into an image completion task offers significant spatio-temporal representation advantages. During the image completion process, the relationships between pixels in different dimensions of the signal image are fully learned, meaning the correlations between different channels and time points in the EEG signals are captured. Through this method, we can obtain and mine the spatio-temporal information of multi-channel EEG signals, providing a novel research approach to address the spatio-temporal representation problem for EEG signals.

Secondly, EEG-DIF provides an effective solution to predict and generate signals from multiple channels with a single model. As shown in **Table 2**, even when predicting 16 signals simultaneously, EEG-DIF exhibits superior prediction performance compared to LSTM-based single-channel prediction method, with a significant improvement in average prediction performance. There are also two signals (F3 and F7) that are less satisfactory by using $R^2$ metric. However, as shown in **Figure 3**, despite the noticeable amplitude changes in the prediction results, EEG-DIF can still accurately predict the signal trends and significant peaks within the next minute. Since These the signal trends and significant peaks are closely related to epileptic seizures [6, 56], it demonstrates that the generated data can provide valuable information for seizure diagnosis. By the way, we suspect that the changes in the amplitude of the predicted signals may result from the rescaling of the min-max normalization recovery process.

Thirdly, as illustrated in **Figure 4** and **Table 3**, our method can directly diagnose epileptic seizures based on the generated future signal data. In other words, it can predict whether seizures will occur in the near future, addressing the challenge of early warning for epileptic seizures. Furthermore, **Figure 4** and **Table 3** turn out that CNN-LSTM is an effective method for EEG signal classification, demonstrating higher accuracy and AUC than other algorithms, thus providing experimental evidence for early seizure warning.

We also indicate the limitations of current method, primarily concerning EEG-DIF's inability to achieve highly accurate future predictions for all 16 different signals, with relatively poor evaluation metrics for some individual signals. We believe this might be related to the order of signal channels in the signal image. Changing the order of the signal channels might improve predictive performance. In future research, we will further explore the relationship between the signal arrangement order in the signal image and the performance of multi-signal forecasting during the image inpainting process.

Moreover, although we have proposed this novel multi-signal forecasting algorithm EEG-DIF, we have not comprehensively explored and evaluated its predictive potential. In this study, we construct and generate predictions for a model using only 16 selected EEG signal channels. Theoretically, this algorithm should generate predictions for any number of channels and any future time length. This is closely related to parameters such as the size of the signal image and the dimensions of the image's noising and inpainting regions. Thus, further study will investigate whether these different parameter settings affect signal prediction performance by incorporating more channel signals into experiments.

Finally, our generative AI model is based on DDIM. Although DDIM is already faster than other generative algorithms like DDPM [47], further research is needed to explore consistency models [62] and other diffusion acceleration algorithms to increase the prediction speed of EEG-DIF for better clinical deployment and application.

In conclusion, EEG-DIF offers innovative ideas and solutions for spatio-temporal representation and future forecasting of multi-channel signals, and early warning of epileptic seizures. Meanwhile, we need continue to explore how to improve EEG-DIF's predictive performance across all channel signals and its inference speed, to develop more efficient multi-channel EEG signals modeling algorithms.

## ACKNOWLEDGMENTS

This work was supported by grants from National Natural Science Foundation of China (62372316), National Science and Technology Major Project (2021YFF1201200), Sichuan Science and Technology Program key project (2024YFHZ0091), and the 1·3·5 Project for Disciplines of Excellence, West China Hospital, Sichuan University (Nos. ZYYC21004).

## Supplementary

The algorithm 1-4 and the code of our algorithm EEG-DIF is located in GitHub: https://github.com/JZK00/EEG-DIF.